\documentclass[12pt]{iopart}
\usepackage{iopams}
\expandafter\let\csname equation*\endcsname\relax
\expandafter\let\csname endequation*\endcsname\relax
\usepackage[fleqn]{amsmath}
\usepackage{amssymb}
\usepackage{amsbsy}
\usepackage{color}
\usepackage{ulem}
\usepackage{graphicx}

\begin{document}

\title{3~m$\times$3~m heterolithic passive resonant gyroscope with cavity length stabilization}

\author{Fenglei Zhang$^{1}$, Kui Liu$^{1, *}$, Zongyang Li$^{1}$, Xiaohua Feng$^{1}$, Ke Li$^{1}$, Yanxia Ye$^{1}$, Yunlong Sun$^{1}$, Leilei He$^{1}$, K. Ulrich Schreiber$^{2}$, Jun Luo$^{3}$, Zehuang Lu$^{1}$, and Jie Zhang$^{1, *}$}

\address{$^{1}${MOE Key Laboratory of Fundamental Physical Quantities Measurements\\
 \& Hubei Key Laboratory of Gravitation and Quantum Physics, PGMF and School of Physics, Huazhong University of Science and Technology, 430074 Wuhan, China}}

\address{$^{2}${Technical University of Munich, Forschungseinrichtung Satellitengeodäsie, \\Geodetic Observatory Wettzell, 93444 Bad Kötzting, Germany}}

\address{$^{3}${TianQin Research Center for Gravitational Physics and School of Physics and Astronomy, \\Sun Yat-sen University (Zhuhai Campus), Zhuhai 519082, China}}

\ead{liukui2007@hust.edu.cn; jie.zhang@mail.hust.edu.cn} \vspace{10pt}
\begin{indented}
\item[] March 3, 2020
\end{indented}

\begin{abstract}

Large-scale high sensitivity laser gyroscopes have important applications for ground-based and space-based gravitational wave detection. We report on the development of a 3~m$\times$3~m heterolithic passive resonant gyroscope (HUST-1) which is installed on the ground of a cave laboratory. We operate the HUST-1 on different longitudinal cavity modes and the rotation sensitivity reaches $1.6\times10^{-9}~$rad/s/$\rm \sqrt{Hz}$ beyond 1~Hz. The drift of the cavity length is one of the major sensitivity limits for our gyroscope in the low frequency regime. By locking cavity length to an ultra-stable reference laser, we achieve a fractional cavity length stability of $5.6\times10^{-9}$~m$/\rm \sqrt{Hz}$ at 0.1~mHz, a four orders of magnitude improvement over the unconstrained cavity in the low frequency regime. We stabilize the cavity length of a large-scale heterolithic passive resonant gyroscope through active feedback and realize long-term operation. The rotation sensitivity reaches $1.7\times10^{-7}$~rad/s/$\sqrt{\rm{Hz}}$ at 0.1~mHz, a three orders of magnitude improvement, which is no longer limited by the cavity length drift in this frequency range. 

\end{abstract}

\vspace{2pc} \noindent{\it Keywords}: passive resonant gyroscope, Earth rotation, gravitational waves

\section{Introduction}\label{sec:intro}

Large-scale laser gyroscopes have experienced rapid development in the past thirty years with markedly improved rotational sensitivity and stability. High-precision large-scale laser gyroscopes have been utilized in many fields, such as rotational seismic research, geodesy, and fundamental physics \cite{RSI2013_Schreiber, QEL2012_Schreiber, RPP1997_Stedman, PRL2011_Schreiber, CRP2014_Virgilio, PRD2011_Bosi}. The large-scale laser gyroscopes are also excellent tilt sensors, especially suitable for the active seismic isolation systems in terrestrial gravitational wave detectors \cite{CQG2010_Virgilio, CQG2016_Korth}. It can also be used to directly measure the length of day (LoD) and the instantaneous orientation fluctuation of the Earth rotational axis. For a space-based gravitational wave detector in the Earth's orbits, such as TianQin project, it is important to obtain the linking information between the celestial reference frame and the terrestrial reference frame. A laser gyroscope that is strapped-down to the earth-body can provide a direct and real-time sensing of the Earth rotation axis and the rotation velocity, which is not available from the space geodetic technique, for example the Very Long Baseline Interferometry \cite{RSI2013_Schreiber}. According to the prediction of general relativity,  the stationary field of a rotating body would have a slight difference from that when it is not rotating, this rotational frame-dragging effect is also known as the Lense-Thirring effect \cite{PZ1918_Lense}. The required sensitivity of the rotating Earth is in the order of $10^{-14}~\rm rad/s$ \cite{PRD2011_Bosi}. The best ring laser gyroscope has already demonstrated a relative resolution of $3.5\times10^{-13}~\rm rad/s$ after $10^5$ s of integration \cite{OL2012_Schreiber, OL2013_Schreiber}. A ring laser gyroscope may be the best instrument of choice for the measurement of Lense-Thirring effect in a terrestrial laboratory \cite{PRD2011_Bosi, IJMPD2010_Virgilio}.

Both sensor types, the active ring laser gyroscopes (RLGs) and passive resonant gyroscopes (PRGs) are based on the same operation principle, namely the Sagnac effect \cite{RPP1967_POST}: 

\begin{equation}
f_{s}=\frac{4A}{\lambda P}\vec{n}\cdot \vec{\Omega}=K_s\vec{n}\cdot \vec{\Omega}.
\label{eq1}
\end{equation}
Here $f_{s}$ is the Sagnac frequency, $A$ and  $P$ denote the area surrounded by the cavity and the length enclosed by the two laser beams, $\lambda$ the laser wavelength, $\vec{n}$ the normal vector of the cavity area, $\vec{\Omega}$ the rotation rate vector, and $K_s$ the scale factor. It is quite clear from Eq. \ref{eq1} that a larger gyro size produces a higher Sagnac frequency and hence a better sensitivity of the instrument. However, on the other hand, a larger size reduces the long-term stability of the geometrical frame, which has a negative effect on the performance of laser gyroscopes in the long-term operation.

In the past, the geometrical frame of laser gyroscopes have been actively stabilized with different methods. Even with a monolithic design using a Zerodur spacer, it is still necessary to apply active length control to improve the long-term stability for the 16 $\rm m^2$ G-ring. For stabilizing the geometry of the heterolithic array of RLGs, the 36~$\rm m^2$ GINGER system is planned to utilize active control via the diagonals of the square cavities \cite{EPJP2017_Virgilio}. The concept was tested on GP2 with a size of 1.96~$\rm m^2$ \cite{CQG2014_Belfi}. 
 
Both PRGs and RLGs are very sensitive to environmental effects, such as temperature, atmospheric pressure, tilt, vibration, etc.. This is especially true for large-scale heterolithic gyroscopes. Unlike the RLGs, the PRGs are operated by locking two counter-propagating laser beams to a ring cavity via two independent servo loops \cite{APL1977_Ezekiel, OL1981_Sanders, JGCD1988_Lorenz, CQG2016_Korth, OL2019_Martynov, OL2019_Liu}. In order to reduce the disturbance induced by the backscattering effect in the locking process, a medium-size PRG (0.56 m$^2$) is operated on adjacent longitudinal cavity modes \cite{CQG2016_Korth}. Based on our previous work on a heterolithic PRG HUST-0 (1 m$^2$), which applies the operation on adjacent longitudinal cavity modes, the long-term performance was found to be mainly limited by the drift of the cavity length \cite{OL2019_Liu}.

In this Letter, we report on a 3~m$\times$3~m large-scale heterolithic PRG (HUST-1) with active cavity length stabilization, using an ultra-stable laser as a length reference. The stability of the geometrical frame can also be improved with the cavity length under control. The successful stabilization of the ring cavity length for a large-scale heterolithic passive resonant gyroscope is one of the key requirements for an improved long-term performance of the gyroscopes. The main development goal of HUST-1 is the support of the space-borne gravitational waves detector ``TianQin'' by linking the celestial reference frame of the satellites to the terrestrial body fixed reference frame with high-time-resolution Earth rotation measurements \cite{CQG2016_Luo, JGeodyn2012_Nilsson}.


\section{Experimental setup}
\label{cavity stabilized laser}

\begin{figure}[!h]
\centering
\includegraphics[width=0.7\textwidth]{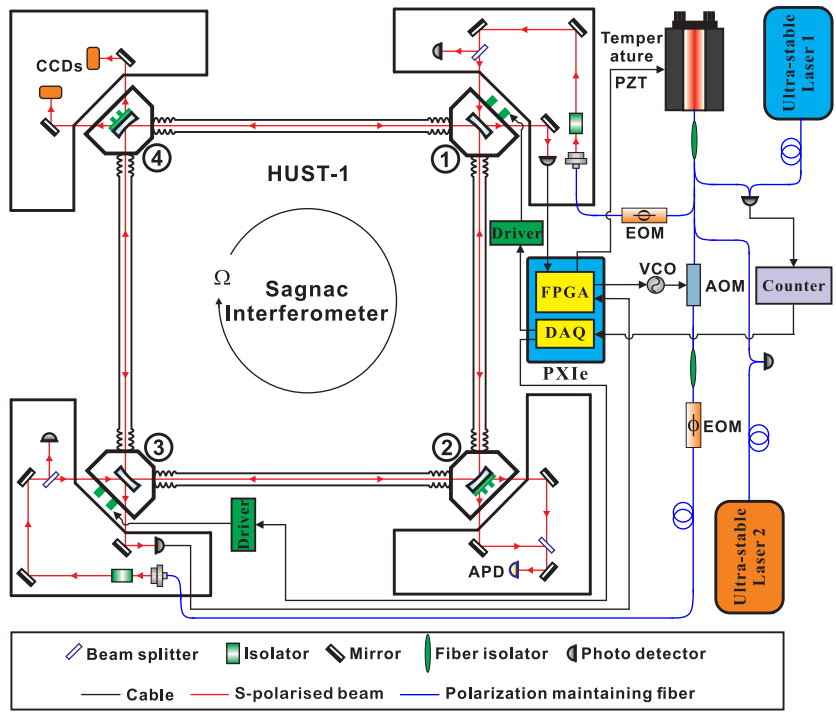}
\caption{Experimental scheme of HUST-1. PZT, piezo-electric transducers of the laser; VCO, voltage controlled oscillator; EOM, electro-optic modulator; AOM, acoustic-optic modulator; FPGA, an Field Programmable Gate Array card (NI PXIe-7858R); DAQ, a data acquisition card (NI PXI-6289); PXIe, PXI express chassis (NI PXIe-1082); APD, avalanche photodiode; CCDs, charge coupled devices;  Ultra-stable Laser 1, an absolute frequency reference ultra-stable laser to stabilize cavity length; Ultra-stable Laser 2, an reference laser to diagnose the ring cavity length noise.} 
\label{setup}
\end{figure}

The square HUST-1 is installed on the ground in a cave lab with a side arm of 3~m, corresponding to a free spectral range (FSR) of 25~MHz. The experimental scheme is shown in Fig. \ref{setup}. Four mirrors with 3~m radius of curvature form the Sagnac interferometer. The mirrors are housed in four stainless steel vacuum chambers, which are placed in four granite corner boxes, respectively. Two of the vacuum chambers can be moved by piezo-electric transducers (PZTs) to control the cavity length. The pressure in the vacuum chamber is as low as $7\times10^{-7}$ Pa to reduce influence of the air flow. The optical Q factor of HUST-1 is measured to be $1.2\times10^{12}$.


The two counter-propagating laser beams are resonant with the ring cavity via two independent servo systems. We use the Pound-Drever-Hall (PDH) technique \cite{APB1983_Drever} to lock the two laser beams to the cavity eigen-modes. The laser beam from a solid state Nd:YAG laser is split into four branches. Two of the branches are used to measure the Earth rotation rate by Sagnac effect. They are phase modulated by two fiber electro-optic modulators (EOM) and injected into the ring cavity on the clockwise (CW) and counterclockwise (CCW) directions, respectively. A fiber acousto-optical modulator (AOM) provides a frequency shift for the CCW laser from the CW loop and acts as an actuator in the locking process. Charge coupled devices (CCDs) at the corner box 4 are used to monitor the the excited optical modes leaking out from the cavity to monitoring the lock status of our PRG. At the corner box 2, we build a Mach-Zehnder heterodyne interferometer and use an avalanche photodetector (APD) to detect the Sagnac beat between the CW and the CCW laser beams. 

In order to realize a long-term undisturbed operation and remote diagnosis of our PRG, we develop a digital locking and control system. The hardware is made up of a Field Programmable Gate Array (FPGA) card (NI PXIe-7858R), inserted into a PXI express chassis (NI PXIe-1082). The FPGA card has 8 high-speed analog differential input channels and 8 high-speed analog output channels, with 16 bits analog-to-digital converter (ADC) and digital-to-analog converter (DAC), operating with a 1 MS/s maximum sampling rate and update rate, which meets the needs of several independent simultaneous looking loops for Sagnac interferometer locking and cavity length stabilization.

As long as the laser is locked to the eigen-mode of the ring cavity, we can record the fluctuation of the cavity length by monitoring the frequency drift of the Nd:YAG gyroscope laser. The frequency drift can be clearly detected by beating the gyroscope laser with an ultra-stable laser 1, taken by a frequency counter and then recorded by a data acquisition (DAQ) card. The ultra-stable laser 1 has a fractional frequency stability better than $1.2\times10^{-14}$ for all integration times from 0.1 s to 10,000 s \cite{RSI2020_Zhang}. This frequency drift serves as an error signal to stabilize the cavity length by locking the PZTs of the vacuum chambers $1\&3$. The last branch of the gyroscope laser output is beating with another independent ultra-stable laser 2 to evaluate the stability of the cavity length under locking conditions.

As long as the primary laser beam tightly follows the ring cavity, the second laser beam only needs to compensate for the Sagnac frequency shift due to the Earth rotation \cite{CQG2016_Korth}. Therefore, the loop gains of the two laser beams are the key parameters of the PRGs. In our work, we set the CW loop as the primary loop. The PDH error signal is acquired and filtered by a second-order infinite impulse response (IIR) low-pass filter with a bandwidth of 100 kHz. Then it is processed by a proportional-integral-integral-differential (PIID) algorithm in FPGA, where the additional integral block is used to increase the loop gain in the low frequency region. The actuators of the primary loop are PZT and crystal temperature of the gyroscope laser. The servo bandwidth of the primary loop is about 30 kHz, limited by the response of the PZT of the Nd:YAG laser, and the loop gain can reach 110 dB at 1 Hz, which is enough for our PRG. 

For the secondary loop, a fiber AOM, driven by a voltage-controlled oscillator (VCO), is used to shift the laser frequency by approximately 75 MHz, which is equivalent to 3 FSRs of the square cavity. We obtain a closed loop locking bandwidth of 30 kHz for the AOM as the actuator. In order to real time evaluate and diagnose the two PDH locking loops, we add two fast Fourier analyzers into the locking program. In addition, auto-locking function is necessary for long-term continuous operation. By monitoring the DC values of the PDH PDs and the APD behind the cavity, we can obtain the locking status of our PRG. If the locking of the primary loop is lost, a frequency-scanning program is launched to search for the TEM$_{00}$ mode. The digital controller implements automatic locking once the TEM$_{00}$ mode is found. All the control codes in the FPGA, the auto-locking and diagnosis programs are realized by LabVIEW.

The operation on different longitudinal cavity modes in our PRG affects the FSR as a result of the fluctuation of the cavity length:
\begin{equation}
\frac{\delta P}{P}=\frac{\delta f_{FSR}}{f_{FSR}}, \label{eq2}
\end{equation}
where  $P$ and $\delta P$ are the 12 m cavity length and its fluctuation, the $f_{FSR}$ denotes the FSR frequency of 25 MHz and $\delta f_{FSR}$ its fluctuation. The variation of the FSR directly influences the Sagnac frequency \cite{OL2019_Liu}:
\begin{equation}
f_s=f_{beat}-3f_{FSR} \label{eq3},
\end{equation}
\begin{equation}
\delta f_{s\_{}FSR}=3f_{FSR}\frac{\delta P}{P}. \label{eq4}
\end{equation}
Here $f_s$ is the Sagnac frequency, $f_{beat}$ is the measured beat frequency from the Mach-Zehnder interferometer which contains the Sagnac frequency $f_s$ and 3$f_{FSR}$ . $\delta f_{s\_{}FSR}$ is the Sagnac frequency fluctuation induced by the variation of FSR frequency, which contributes the majority of frequency noise of Sagnac frequency $f_s$ of our PRG, especially in the low frequency region.

\section{Experimental results and discussion}
\label{results} 

\begin{figure}[!h]
\centering
\includegraphics[width=0.7\textwidth]{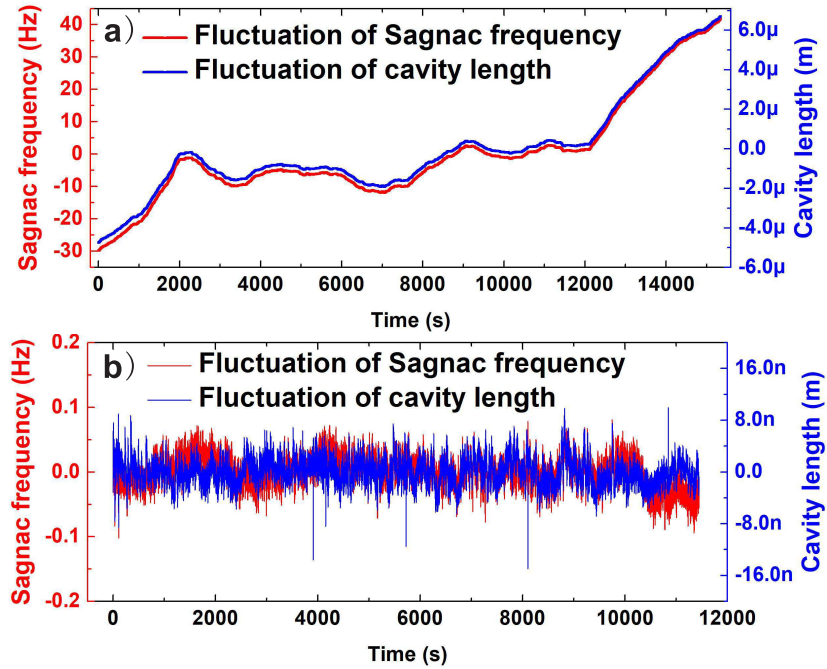}
\caption{a) The fluctuations of the Sagnac frequency and the cavity length are shown with a red curve and a blue curve in an unconstrained ring cavity length. b) The residual fluctuations of the Sagnac frequency and the cavity length are shown with a red curve and a blue curve with a stabilized ring cavity length.}
\label{cavitylength}
\end{figure}

\begin{figure}[!h]
\centering
\includegraphics[width=0.65\textwidth]{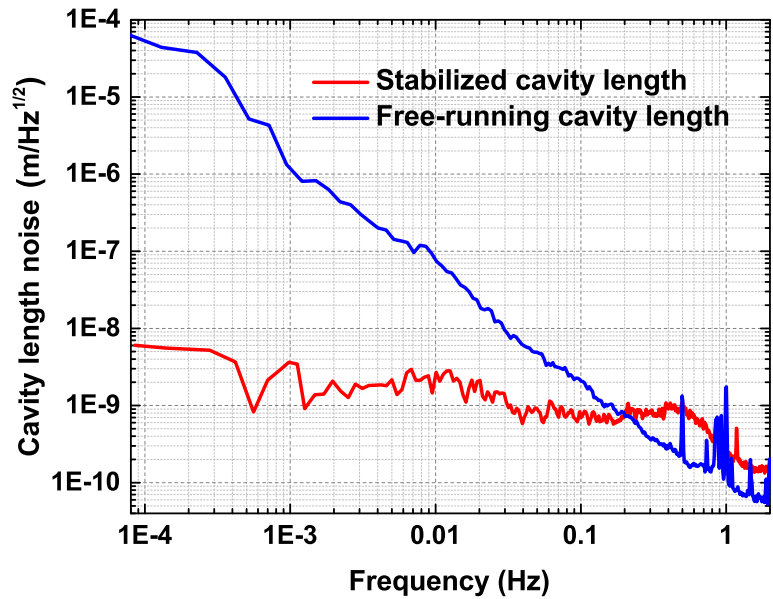}
\caption{The linear power spectral density of the cavity length fluctuation. The blue and red curve represent the cavity length fluctuation when the cavity length is free running and locked, respectively.}
\label{PSDcavitylength}
\end{figure}

As shown in Fig. \ref{cavitylength} a), we measure the fluctuation of the Sagnac frequency in an unconstrained ring cavity (red curve). We also monitor the cavity length fluctuation at the same time (blue curve). The fluctuation of the Sagnac frequency is affected by the cavity length fluctuation. In order to stabilize the cavity length, we use the ultra-stable laser 1 as an absolute length reference. When the laser frequency of the primary loop resonates with the ring cavity, the fluctuation of the cavity length can be expressed by the frequency variation of the CW laser beam. By monitoring the frequency difference between the CW laser beam and the reference laser, an error signal due to the fluctuation of the cavity length can be recorded by a frequency counter. The error signal is processed by a digital controller (PXI-6289) in the PXIe chassis, and is fed back to the two PZTs under the vacuum chambers via PZT drivers. In order to maintain the geometric symmetry of the square cavity, two PZTs are driven on opposite corners to push two vacuum chambers synchronously for the compensation of the drift of the cavity length. The ultra-stale laser 2, which has a frequency stability of $8\times10^{-16}$ at 1 s and a drift rate of 3 kHz$/$day \cite{RSI2016_Zhang_Shi, OL2018_Zeng}, is used to estimate the stability of the cavity length, when the cavity length is locked to the ultra-stable laser 1. The residual cavity length fluctuation with cavity length locked is shown in the blue curve in Fig. \ref{cavitylength} b), and the Sagnac frequency is measured simultaneously, shown in the red curve. The linear power spectral density of the 12 m cavity length fluctuation is shown in Fig. \ref{PSDcavitylength}. The displacement noise of the 12 m cavity length (red curve) is about $5.6\times10^{-9}$ m$/\rm \sqrt Hz$ at 0.1 mHz and it is improved by four orders of magnitude over the unconstrained cavity. In the frequency range from 0.5 Hz to 1 mHz, the residual cavity length noise acts as the white noise. There is a small servo bump at about 0.6 Hz, indicating the closed loop bandwidth of the cavity length locking loop. Above 0.6 Hz, higher noise level is caused by the output voltage noise of the PXI-6289 card.

\begin{figure}[!h]
\centering
\includegraphics[width=0.7\textwidth]{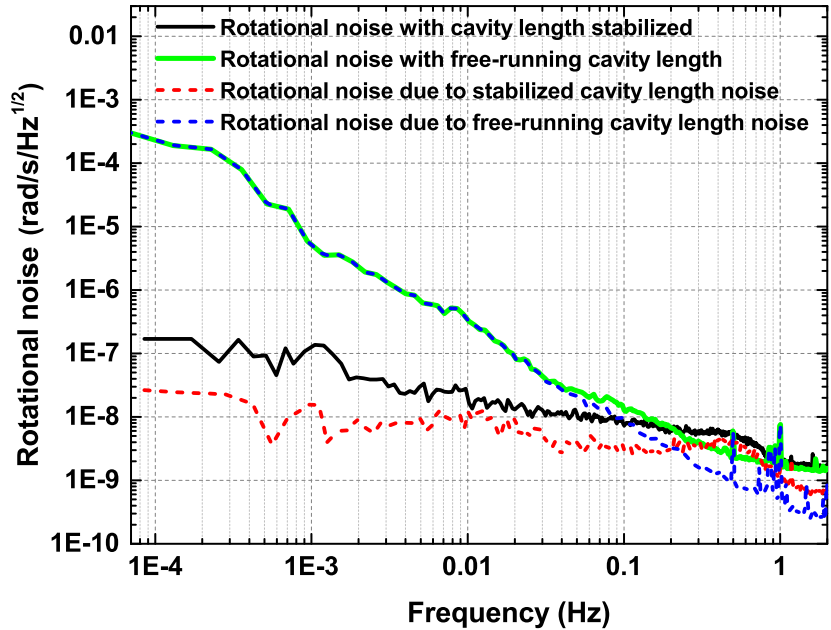}
\caption{The linear power spectral density of the rotation sensitivity of our PRG. The blue and red dotted curves show the free and stabilized cavity length fluctuations and their contribution to the rotation sensitivity of our PRG. The measured rotation sensitivities with the free and stabilized cavity length are shown with the green and black solid curve.}
\label{psdsagnacfrequency}
\end{figure}

The performance of the HUST-1 is shown as linear power spectral density in Fig. \ref{psdsagnacfrequency}. The rotation sensitivity contributions due to the cavity length fluctuations are also shown in the same figure for comparison. We can see that rotational noise of $1.7\times10^{-8}$ rad/s/$\rm \sqrt{Hz}$ is converted from the 12 m cavity length noise with the cavity length stabilized (red dotted curve). The green solid curve represents the measured rotational noise when the cavity length is not locked, which is nearly identical with the equivalent rotational noise due to the unconstrained cavity length noise (blue dotted curve). The rotation sensitivity of our PRG with cavity length stabilized (black solid curve) is $1.6\times10^{-9}$ rad/s/$\rm \sqrt{Hz}$ at 1 Hz and $1.7\times10^{-7}$ rad/s/$\rm \sqrt{Hz}$ at 0.1 mHz. At this time, the HUST-1 is no longer limited by the cavity length drift at this frequency region. Other possible rotational noises in our system in the low frequency range include the power fluctuation, polarization variaton of the two injected lasers, environmental noise of the cave lab, and the residual amplitude modulation (RAM) in the PDH locking process, etc.. More research will be done in the future to suppress these contributions.

\section{Conclusion}\label{conclusion} 

In conclusion, we have developed a 3 m$\times$3 m large-scale heterolithic PRG HUST-1, which has a rotation sensitivity of $1.6\times10^{-9}$ rad/s/$\rm \sqrt{Hz}$ beyond 1 Hz by operating on different longitudinal cavity modes. Digital controllers are implemented with a FPGA card for the PDH locking scheme, and meet the requirement of auto-locking and long-term operation of HUST-1. We accomplish the stabilization of the 12 m cavity length by locking cavity length to a reference laser, and the stability of the cavity length is $5.6$ nm$/\rm \sqrt Hz$ at 0.1 mHz with an improvement of four orders of magnitude over the unconstrained cavity. We achieve a rotation sensitivity of $1.7\times10^{-7}$ rad/s/$\rm \sqrt{Hz}$ at 0.1 mHz with the cavity length stabilized. To our knowledge, so far this is the best result achieved among all large-scale PRGs. This indicates that large-scale heterolithic PRGs have great potential to be operated with high sensitivity and long-term stability, simultaneously. Limited by the high load PZTs under the vacuum chambers, the bandwidth of the cavity length locking loop is not high enough. In the future, we will use two fast PZTs to directly push the cavity mirrors to increase the servo loop bandwidth. In addition, passive/active temperature isolation is also quite necessary for the long-term operation. Moreover, new methods should be developed to suppress the RAM effect. With these improvements in place, we believe much better sensitivity and long-term stability can be expected for HUST-1.

\section{Acknowledgments} 
 The project is partially supported by the National Key R\&D Program of China (Grant No. 2017YFA0304400), the Key-Area Research and Development Program of GuangDong Province (Grant No. 2019B030330001), the National Natural Science Foundation of China (Grant No. 91536116, 91336213, and 11774108), and China Postdoctoral Science Foundation (Grant No. 2018M642807).

\section*{References}

\normalem

\end{document}